\documentclass[twocolumn,showpacs,preprintnumbers,amsmath,amssymb]{revtex4}

\usepackage{graphicx}
\usepackage{dcolumn}
\usepackage{bm}

\begin{document}


\title{Superconductors with Topological Order}
\author{M. Cristina Diamantini}
\email{cristina.diamantini@pg.infn.it}
\affiliation{%
INFN and Dipartimento di Fisica, University of Perugia, via A. Pascoli, I-06100 Perugia, Italy
}%

\author{Pasquale Sodano}
\email{pasquale.sodano@pg.infn.it}
\affiliation{%
INFN and Dipartimento di Fisica, University of Perugia, via A. Pascoli, I-06100 Perugia, Italy
}%

\author{Carlo A. Trugenberger}
\email{ca.trugenberger@InfoCodex.com}
\affiliation{%
InfoCodex S.A., av. Louis-Casai 18, CH-1209 Geneva, Switzerland
}%


\date{\today}

\begin{abstract}
We propose a mechanism of superconductivity in which the order of the ground state does
not arise from the usual Landau mechanism of spontaneous symmetry breaking but is rather of 
topological origin. The low-energy effective theory is formulated in terms of 
emerging gauge fields rather than a local order parameter and the ground state is degenerate
on topologically non-trivial manifolds. The simplest example of
this mechanism of superconductivty is concretely realized as global superconductivty 
in Josephson junction arrays. 

\end{abstract}

\maketitle
The discovery  of the fractional quantum Hall \cite{dqhl} effect has revealed the existence of a new state of 
matter characterized by a new type of order: topological order \cite{wen1}. Topological order is a particular type of quantum order
describing zero-temperature properties of a ground state with a gap for all excitations. 
Its hallmark are the degeneracy of the ground state on manifolds
with non-trivial topology, and excitations with fractional spin and statistics, called anyons \cite{wilczek}. The long-distance properties
of these topological fluids are described by Chern-Simons field theories \cite{wen2} with compact gauge group, which break P-and 
T-invariance. Other examples of P-and T-breaking topological fluids are given by chiral spin liquids \cite{bob}.

After Laughlin's discovery of topological quantum fluids, it was conjectured that
a similar mechanism, based on anyon condensation, could be at the origin of high-$T_c$ 
superconductivity \cite{wilczek}. Unfortunately, there is no evidence of the associated
broken P- and T-invariance in the high-$T_c$ materials. 

Here we propose a superconductivity mechanism which is based on a topologically ordered 
ground state rather than on the usual Landau mechanism of spontaneous symmetry breaking.Contrary to anyon superconductivity it works in any dimension
and it preserves P-and T-invariance. In particular we will discuss the low-energy effective field theory, what would be the Landau-Ginzburg
formulation for conventional superconductors.

Topologically ordered superconductors have a long-distance hydrodynamic action which 
can be entirely formulated in terms of generalized compact gauge fields, the dominant 
term being the topological BF action.

BF theories are  topological theories that can be defined on manifolds $M_{d+1}$ of any 
dimension (here d is the number of spatial dimensions) and play a crucial role in models of two-dimensional gravity
\cite{marito}.
In \cite{ccp1} we have shown that the BF term also plays a crucial role in the physics of Josephson junction arrays.
 
The BF term \cite{birmi} is the wedge product of a p-form B  and the curvature $d A$ of a (d-p) form A:

\begin{equation}
S_{BF} = {k \over 2 \pi} \int_{M_{d+1}}  B_p \wedge d A_{d-p}\ ,
\nonumber
\end{equation}
where $k$ is a dimensionless coupling constant.
This can also be written as
\begin{equation}
S_{BF} = {k \over 2 \pi} \int_{M_{d+1}}  A_{d-p} \wedge d B_p\ .
\label{sbf1}
\end{equation}
The integration by parts does not imply any surface term since we will concentrate on compact spatial manifolds without
boundaries and we require that the fields go to pure gauge configuration at infinity in the time direction.
Indeed this action has a generalized Abelian gauge symmetry under the transformation 
\begin{equation}
B \rightarrow B + \eta \ ,
\nonumber
\end{equation}
where $\eta$ is a closed p form: $d \eta = 0$.
Gauge transformations:

\begin{equation}
A \rightarrow A + \xi \ ,
\nonumber
\end{equation}
with $\xi$ a closed (d-p) form instead, change the action by a surface term. This, however vanishes with the boundary conditions we have
chosen.

Here we will be interested in the special case where $A_1$ is a 1-form and, correspondingly, $B_{d-1}$ is a (d-1)-form: 
\begin{equation}
S_{BF} = {k \over 2 \pi} \int_{M_{d+1}} A_1 \wedge d B_{d-1} \ .
\label{sbf}
\end{equation}
In the special case of (3+1) dimensions,  B is the well-known Kalb-Ramond tensor field $B_{\mu\nu}$ \cite{kalb}. 

In the application to superconductivity, the conserved current $j_1 = * dB_{d-1}$ 
represents the charge fluctuations, while the generalized current 
$j_{d-1} = * dA_1$ describes the conserved fluctuations of (d-2)-dimensional 
vortex lines. As a consequence,
the form $B_{d-1}$ must be considered as a pseudo-tensor, while $A_1$ is a vector, as
usual. The BF coupling is thus P- and T-invariant. 

The low-energy effective theory of the superconductor can be entirely expressed in terms
of the generalized gauge fields $A_1$ and $B_{d-1}$. The dominant term at long distances is
the BF term; the next terms in the derivative expansion of the effective theory are the
kinetic terms for the two gauge fields (for simplicity of 
presentation we shall assume relativistic invariance), giving:

\begin{eqnarray}
S_{TM} &=  \int_{M_{d+1}} { - 1 \over 2 e^2} d A_1 \wedge * d A_1 +  {k \over 2 \pi} A_1 \wedge d B_{d-1} \nonumber \\
&+ {(- 1)^{d-1} \over 2 g^2} d B_{d-1}\wedge * d B_{d-1}\ ,
\label{topmas}
\end{eqnarray}  
where $e^2$ and $g^2$ are coupling constants of dimension $m^{-d+3}$ and $m^{d-1}$ respectively.

The BF-term is the generalization to any number of dimensions of the Chern-Simons mechanism for the topological mass
\cite{jackiw}. To see this let us now compute the equation of motion for the two forms $A$ and $B$:

\begin{equation}
{1 \over g^2} d*d B_{d-1} = {k \over 2 \pi} d A_1 \ ,
\label{eqmo1}
\end{equation} 
and
\begin{equation}
{1 \over  e^2} d*d A_1 = {k \over 2 \pi} d B_{d-1} \ .
\label{eqmo2}
\end{equation}  
Applying $d*$ on both sides of (\ref{eqmo1}) and (\ref{eqmo2}) we obtain 

\begin{eqnarray}
 &d*d* d A_1 - {k e^2\over 2 \pi} d*d B_{d-1} = 0 \ , \nonumber \\
 &d*d* d B_{d-1} - {k g^2\over 2 \pi} d*d A_1 = 0 \ .
\label{is}
\end{eqnarray} 
The expression $*d*$ is proportional  to $\delta$, the adjoint of the exterior derivative \cite{eh}.
Substituing  $d*d B_{d-1}$  and 
$d*d A_1$ in (\ref{is}) with the expression coming from (\ref{eqmo1}) and (\ref{eqmo2}) we obtain

\begin{eqnarray}
&\left( \Delta + m^2 \right) d A_1 = 0 \ , \nonumber \\
&\left( \Delta + m^2 \right) d B_{d-1} = 0 \ ,
\label{topmas1}
\end{eqnarray}                                                                                                                       
where $\Delta = d \delta$ (when acting on an exact form) and  $m = {k e g \over 2 \pi}$ is the topological mass.
This topological mass plays the role of the gap characterizing  the superconducting ground state.
Note that the gap arises here from a topological mechanism and not from a local order parameter acquiring a vacuum expectation value.
Equations (\ref{eqmo1}) and (\ref{eqmo2}) tell us that charges are sources for vortex 
line currents encircling them and viceversa. This is the coupling between charges and
vortices at the origin of the gap. A related mechanism for topological mass generation in (3+1)-dimensional gauge theories is the
generalization of the Schwinger mechanism proposed in \cite{dvali}.

Let us now consider the special case of (2+1) dimensions (d=2). In this case also
$B$ becomes a (pseudo-vector) 1-form and, correspondingly the BF term reduces to a
mixed Chern-Simons term. This can be diagonalized by a transformation $A={a+b\over 2}$, 
$B=a-b$, giving
\begin{equation}
S_{BF}(d=2) = {k\over 4\pi} \ \int a \wedge da - {k\over 4\pi} \ \int b \wedge db \ .
\label{newa}
\end{equation}
The result is a doubled Chern-Simons model for gauge fields of opposite chirality. This action, including its non-Abelian generalization with
kinetic terms was first considered in \cite{roman}. It is
the simplest example of the class of P- and T-invariant topological phases of strongly
correlated (2+1)-dimensional electron systems considered in \cite{freedman}. Indeed, the BF
term is the natural generalization of such doubled Chern-Simons models to any dimension.
Doubled (or mixed) Chern-Simons models are thus particular examples in two spatial
dimensions of a wider class of P- and T-invariant topological fluids that have a 
superconducting phase. These fluids are described by the topological BF theory
with compact support for both gauge fields.

Topological BF models provide also a generalization of anyons to arbitrary dimensions. While in (2+1) dimensions fractional statistics arises
from the representations of the braid group, encoding the exchange of particles, in (3+1) dimensions it arises from the adiabatic transport of
particles around vortex strings and, in  (d+1) dimensions, from the motion of an hypersurface $\Sigma_h$ around  another hypersurface $\Sigma_{d-h}$.
The relevant group in this case is the motion group and the statistical parameter is given by ${2 \pi \over k} h (d-h)$, where
$k$ is the BF coupling constant \cite{szab}.

Let us now illustrate the mechanism of superconductivity. To this end we shall from now on  consider only rational $k = {k_1 \over k_2}$ with
$k_i$ integers, and specialize to manifolds $M_{d+1} = M_d \times R_1$, whith $R_1$ representing the time direction. 

The compactness of the gauge fields allows for the presence of topological defects,
both electric and magnetic. The electric topological defects  couple to the form
$A_1$ and are string-like objects  described by a singular closed 1-form $Q_1$. Magnetic topological 
defects couple to the form
$B_{d-1}$ and are closed (d-1)-branes described  by a singular (d-1)-dimensional form $\Omega_{d-1}$. These forms
represent the singular parts of the field strenghts $dA_1$ and $dB_{d-1}$, allowed by the compactness of the gauge
symmetries \cite{polbo}, and are such that the  integral of their Hodge dual  over any  hypersurface of dimensions d and 2, respectively, is $2 \pi$ times
an integer as can be easily derived using a lattice regularization. Contrary to the currents $j_1$ and $j_{d-1}$, which represent charge- and
vortex-density waves, the topological defects describe localized charges and vortices. In the effective theory these have structure on the
scale of the ultraviolet cutoff.

We will not discuss here the conditions for the condensation of topological defects, but we will show, instead that the phase 
of electric condensation describes  a superconducting phase in any dimension.
A detailed analysis would require the use of an ultraviolet regularization.
Here we will present a formal derivation implying the ultraviolet regularization. A detailed derivation on the lattice
 will follow in a forthcoming publication \cite{ccp2}.

In the phase in which electric topological defects condense (while magnetic ones are dilute) the partition function requires 
a formal sum also over the form $Q_1$

\begin{eqnarray}
&Z = \int {\cal D}A {\cal D}B {\cal D}Q \nonumber \\ 
&\exp \left[ i {k \over 2 \pi}\int_{M_{d+1}} \left( A_1 \wedge d B_{d-1} + A_1 \wedge *Q_1 \right) \right] \ .
\label{pfbf}
\end{eqnarray}
Let us now compute the expectation value of the 't Hooft  operator, 
$\langle L_H \rangle$, which represents
the amplitude for creating and separating a pair of vortices with fluxes $\pm \phi$:

\begin{eqnarray}
&\langle L_H \rangle = {1 \over Z} \int {\cal D}A {\cal D}B {\cal D}Q \nonumber \\ 
&\exp \left[ i {k \over 2 \pi}\int_{M_{d+1}} \left( A_1 \wedge d B_{d-1} + A_1 \wedge *Q_1 \right) \right. \nonumber \\
&+ \left. i {k \over 2 \pi} \phi \int_{S_{d-1}} B_{d-1} \right]\ .
\label{pfbh}
\end{eqnarray}
Using Stokes' theorem we can rewrite

\begin{equation}
\int_{S_{d-1}} B_{d-1}  = \int_{S_d} d B_{d-1} \ ,
\label{sto}
\end{equation}
where the surface $S_d$ is such that $\partial S_d \equiv  S_{d-1}$ and represents a compact orientable surface on $M_d$.
Inserting (\ref{sto}) in (\ref{pfbh}) and integrating over the field $A$ we obtain:

\begin{eqnarray}
\langle L_H \rangle  &  \propto \int  {\cal D}B {\cal D}Q \ \delta \left( d B_{d-1} + *Q_1 \right) \nonumber \\ 
&\exp \left[  i{k \over 2 \pi}  \phi \int_{S_d} d B_{d-1} \right]\ .
\end{eqnarray}
Integrating over B gives then:

\begin{equation}
\langle L_H \rangle = \propto \int {\cal D}Q\ \exp \left[ - i{k \over 2 \pi} \phi \int_{S_d} *Q_1 \right] \ .
\label{vev}
\end{equation}
The Poisson summation formula implies finally that the 't Hooft loop expectation value
vanishes for all flux strengths $\phi $ different from 
\begin{equation}
{\phi \over k_2} = {2\pi \over k_1} \ n \qquad \qquad n \in N \ .
\label{newb}
\end{equation}
This is nothing else than the Meissner effect, illustrating that the electric condensation
phase is superconducting. Indeed, the electric condensate carries $k_1$ fundamental
charges of unit $1/k_2$ as is evident from (\ref{pfbf}), and
correspondingly vortices must carry an integer multiple of the fundamental fluxon
$2\pi /({k_1/k_2})$. All other vorticities are confined: in this purely 
topological long-distance theory the confining force is infinite; including 
the higher order kinetic terms (\ref{topmas}) and the UV cutoff one would recover a generalized area law. 

Another way to see this is to compute the current induced by an external electromagnetic
field $A_{\rm ext}$. The corresponding coupling is $\int_{M_{d+1}} A_{\rm ext}
\wedge \left( * j_1 + *Q_1 \right)$ $\propto $ $\int_{M_{d+1}} A_{\rm ext}
\wedge \left( dB_{d-1} + *Q_1 \right)$. Since $A_{\rm ext}$ can be entirely reabsorbed in a redefinition
of the gauge field $A_1$, the induced current vanishes identically, $j_{\rm ind} =0$.
This is just the London equation in the limit of zero penetration depth. Including the
higher-order kinetic terms for the gauge fields and the UV cutoff one would again recover the standard form
of the London equation. 

Associated with the confinement of vortices there is a breakdown of the original U(1)
matter symmetry under transformations $A_1 \to A_1 + d\lambda$. To see this let us consider 
the effect of such a transformation on the partition function (\ref{pfbf}) with an
electric condensate. Upon integration by parts, the exponential of the action acquires 
a multiplicative factor
\begin{equation}
{\rm exp} \ i {k_1\over 2\pi k_2} \left( \int_{M_d , t=+\infty} \lambda \wedge * Q_1
-\int_{M_d , t=-\infty} \lambda \wedge * Q_1 \right) \ .
\label{newc}
\end{equation}
Assuming a constant $\lambda $, we see that the only values for which the partition
function remains invariant are 
\begin{equation}
\lambda = 2\pi \ n \ {k_2 \over k_1}\ , \qquad \qquad n=1 \dots k_1 \ ,
\label{newd}
\end{equation}
which shows that the global symmetry is broken from U(1) to $Z_{k_1}$. Note that this
is not the usual Landau mechanism ofspontaneous symmetry breaking. Indeed, there is
no local order parameter and the order is characterized rather by 
the expectation value of non-local, topological operators.

The hallmark of topological order is the degeneracy of the ground state on manifolds with non-trivial topology as shown by Wen \cite{wen1}. 
In (2+1) dimensions the degeneracy for the mixed Chern-Simons term was proven in \cite{hoso} for the case of integer coefficient $k$ 
of the Chern-Simons term.

The degeneracy of the ground state of the BF theory on a manifold with non-trivial topology  was proven in \cite{gord} in (3+1) dimensions.
This result can be generalized to compact topological BF models in any number of dimensions \cite{szab}.
Consider the model (\ref{sbf1}) with $k = {k_1 \over k_2}$ on a manifold $M_d \times R_1$, 
with $M_d$ a compact, path-connected , orientable d-dimensional manifold without boundaries. The degeneracy of the ground state is expressed
in terms of the intersection matrix $M_{mn}$ \cite{bott} with $m,n = 1....N_p$ and $N_p$  the rank of the matrix, 
between  p-cycles and  (d-p)-cycles. $N_p$ corresponds to the number of generators of the two homology groups
$H_p(M_d)$ and $H_{d-p}(M_d)$ and is essentially the number of non-trivial cycles on 
the manifold $M_d$. The degeneracy of the ground state is given by $|k_1 k_2 M|^{N_p}$,
where $M$ is the integer-valued determinant of the linking matrix. In our case p = (d-1) and the degeneracy reduces to

\begin{equation}
|k_1 k_2 M|^{N_{d-1}} \ .
\label{deged}
\end{equation}

In this paper we have derived a superconductivity mechanism which is not based on the 
usual Landau theory of spontaneous symmetry breaking. Our considerations here focused on 
the low-energy effective theory in order to expose
the physical basis of the topological superconducivity mechanism. It is however crucial
to stress that the simplest example (k=1) of 
this type of topological superconductivity is concretely realized as the global 
superconductivity mechanism in 
planar Josephson junction arrays, as we have shown in \cite{ccp1}. Naturally, it
would be most interesting to find examples of microscopic models realizing this
superconductivity mechanism with more complex degeneracy patterns. The exsistence of such non-conventional superconductors
is also supported by purely algebraic considerations \cite{carlo}

\end{document}